\documentclass[twocolumn,twoside,pra,aps,superscriptaddress,showpacs]{revtex4}
\usepackage{amsmath,mathrsfs,amsbsy,color,graphicx,bm,amsthm,amsfonts}
\usepackage{units}
\usepackage{bbm}

\begin{document}

\title{Control of single-photon transport in a one-dimensional waveguide by
another single photon}
\author{Wei-Bin Yan}
\affiliation{Liaoning Key Lab of Optoelectronic Films \& Materials,
School of Physics and Materials Engineering, Dalian Nationalities
University, Dalian, 116600, China}
\affiliation{Beijing National Laboratory for Condensed Matter Physics, Institute of
Physics, Chinese Academy of Sciences, Beijing 100190, China}
\author{Heng Fan}
\email{hfan@iphy.ac.cn}
\affiliation{Beijing National Laboratory for Condensed Matter Physics, Institute of
Physics, Chinese Academy of Sciences, Beijing 100190, China}

\begin{abstract}
We study the controllable single-photon transport in a one-dimensional (1D)
waveguide with nonlinear dispersion relation coupled to a three-level
emitter in cascade configuration. An extra cavity field was introduced to
drive one of the level transitions of the emitter. In the resonance case,
when the extra cavity does not contain photons, the input single photon will
be reflected, and when the cavity contains one photon, the full transmission
of the input single photon can be obtained. In the off-resonance case, the
single-photon transport can also be controlled by the parameters of the
cavity. Therefore, we have shown that the single-photon transport can be
controlled by an extra cavity field.
\end{abstract}

\pacs{03.67.Hk, 03.65.-w}
\maketitle

\section{Introduction}

Single photons are considered as one of the most suitable carriers for
quantum information. It is important to control the single-photon transport
in quantum information processing. Recently, controllable single-photon
transport in 1D waveguide with linear and nonlinear dispersion relations has
been extensively investigated both theoretically \cite%
{Shen2005ol,Lan,Shen2007pra,Roy2010prb,royb,Fan2012prl,kocabas,Shenjt,fanshen,Paolo,busch,Shit,Zhengprl,Zhengprl1,Zhengprl2,Liao,Witthaut2010njp,liqiong,zhour,Gongzr,Liaojq,Weilf1,Cheng,Wangzh,brad,Yanwb,Weilf,Hart}
and experimentally\cite{circuit,1,2,Akimov,Bajcs,
Babinec,Claudon,Bleuse,Laucht}. In a 1D waveguide, the photons are confined
to propagating only forward or backward in 1D space. By coupling an emitter
to the waveguide, the strong photon-emitter interaction can be obtained and
the photon transport in the 1D space can be affected by the interaction. It
is known that for a 1D waveguide coupled to a two-level emitter \cite%
{Shen2005ol,Lan}, the injected single photon will be completely reflected in
the resonance case due to the interference between the wave function of the
input photon and the spontaneously emitted photon. The single-photon
transport in a 1D waveguide coupled to a multi-level emitter \cite%
{Witthaut2010njp,Gongzr} has also been studied. Compared to a two-level
system, the multi-level system provides more controllable parameters. For
instance, when a 1D waveguide is coupled to a $\Lambda $-type emitter, a
strong pulse was employed to drive one of the atomic transitions. Since the
single photon transport can be controlled by the extra pulse, the
all-optical device can be achieved. However, most of the controls need
strong pulse containing many photons or rely on other parameters. It is
interesting to study the controllable single-photon transport by another
photon without the classical field.

In this paper, we propose a scheme to study the control of single-photon
transport in a 1D waveguide with nonlinear dispersion relation at the
single-photon level. In our scheme, the single-photon transport can be
controlled by a cavity field. When the control cavity field is in the vacuum
state, our scheme can be considered as a 1D waveguide coupled to a two-level
system. Therefore, the input single photon will be reflected in the
resonance case. When the control cavity contains one photon, our scheme
becomes a 1D waveguide coupled to a three-level system with cascade
configuration. The full transmission of the input single photon can be
obtained in the resonance case. It is necessary to note that the control of
single-photon transport by another photon without classical field in 1D
waveguide with linear dispersion relation has been studied in Ref. \cite%
{Hart}. We also study the single-photon transport controlled by the
parameters of the control cavity, such as the coupling strength to the
emitter, photon number and resonant frequency. By the way, the quantum
control of single-photon transport in 1D waveguide with linear dispersion
relation by cavity-emitter coupling strength has been studied in Ref. \cite%
{Weilf}. Our scheme does not contain inelastic scattering. Our scheme can be
considered as a 1D waveguide with nonlinear dispersion relation coupled to a
V-type atom in the dressed-state representation. Some outcomes can be
understood better than in the bear-state representation.

\section{Model and Hamiltonian}

The schematic diagram of the considered system is shown in Fig. 1. The 1D
waveguide is composed by a coupled-cavity array with a very large number of
single-mode cavities and one two-mode cavity. The cavity modes of the 1D
coupled cavities are represented by the annihilation operators $a_{j}$, with
$j$ the label of site. Here we take the site of the two-mode cavity $0$.
Hence, one mode of the two-mode cavity, which is coupled to the
nearest-neighbor single-mode cavities, is represented by the annihilation
operators $a_{0}$. We represent the other mode of the two-mode cavity by the
annihilation operator $b$. For simplicity, we assume that the cavity modes $%
a_{j}$ have the same resonant frequency $\omega _{a}$. The cavity mode $b$\
has the resonant frequency $\omega _{b}$.\ A three-level system in cascade
configuration is doped in the two-mode cavity. The three-level system can be
a real atom or a manual atom-like object. It will be mentioned as an atom
below. We represent atomic states $\left| 1\right\rangle $, $\left|
2\right\rangle $ and $\left| 3\right\rangle $, with frequencies $\omega _{1}$%
, $\omega _{2}$ and $\omega _{3}$, respectively. We choose the ground-state
energy for reference and ,hence, take $\omega _{1}$ zero. The level
transitions between $\left| 1\right\rangle \leftrightarrow \left|
2\right\rangle $ and $\left| 2\right\rangle \leftrightarrow \left|
3\right\rangle $ are coupled to the modes $a_{0}$ and $b$ with strengths $%
g_{a}$ and $g_{b}$, respectively. In the rotating-wave approximation, the
system Hamiltonian has the form of%
\begin{equation}
H=H_{C}+H_{A}+H_{I}\text{,}  \label{Hamiltonian}
\end{equation}%
with%
\begin{eqnarray*}
H_{C} &=&\omega _{a}a_{j}^{\dagger }a_{j}+\omega _{b}b^{\dagger }b-\xi
\sum_{j}(a_{j+1}^{\dagger }a_{j}+h.c.)\text{,} \\
H_{A} &=&\omega _{2}\sigma ^{22}+\omega _{3}\sigma ^{33}\text{,} \\
H_{I} &=&g_{a}\sigma ^{21}a_{0}+g_{b}\sigma ^{32}b+h.c.\text{.}
\end{eqnarray*}%
Here we have taken $\hbar =1$. The Hamiltonian $H_{C}$ denotes the cavity
photons, $H_{A}$ is the atomic free Hamiltonian and $H_{I}$ describes the
interaction of the atom with the two-mode cavity. The sum part of $H_{C}$
describes the hopping of the $a_{j}$ mode photons to the nearest-neighbor
cavities with strength $\xi $. Here we have assumed that all the hopping
strengths are equal. By introducing the Fourier transform $a_{k}=\frac{1}{%
\sqrt{N}}\sum_{k}e^{ikj}$ and taking the distances between two
nearest-neighbor cavities unit, the term $\omega _{a}a_{j}^{\dagger
}a_{j}-\xi \sum_{j}(a_{j+1}^{\dagger }a_{j}+h.c.)$ can be diagonalized as $%
\sum_{k}\Omega _{k}a_{k}^{\dagger }a_{k}$, with $\Omega _{k}=\omega
_{a}-2\xi \cos k$ and $N$ being the cavity number. This implies the
nonlinear dispersion relation of the 1D coupled-cavity waveguide.

\begin{figure}[t]
\includegraphics*[width=8cm, height=4cm]{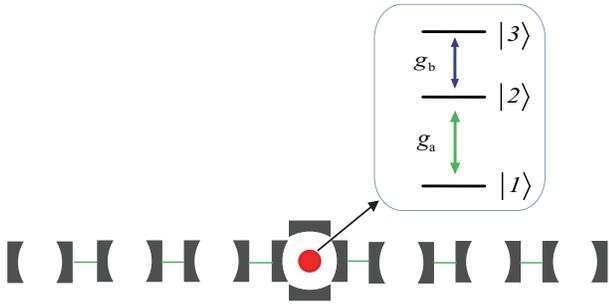}
\caption{Schematic diagram of the single-photon transport in a
one-dimensional coupled-resonator waveguide. A three-level system with
cascade configuration is coupled to the waveguide.}
\end{figure}

\section{Control of single-photon transport}

We assume that, initially, a photon is injected into the waveguide from the
left side, the atom is in its ground state $\left| 1\right\rangle $, and
that the two-mode cavitiy contains $n$ $b$-mode photons. We note that the
value of $n$ can not be very large, i. e. $g_{b}\sqrt{n}\ll \{\omega _{2}$, $%
\omega _{3}\}$. This because we have emploied the rotating-wave
approximation in Hamiltonian (1). The input single photon will transport
along the waveguide and be scattered due to the atom-cavity interaction.
Obviously, when $n=0$, the atom will absorb the injected photon, meanwhile
making a transition from the level $\left| 1\right\rangle $ to $\left|
2\right\rangle $, and then reemit a photon, meanwhile making a transition
from the level $\left| 2\right\rangle $ to $\left| 1\right\rangle $. In this
case, the atomic level $\left| 3\right\rangle $ never participates in the
dynamic process because the atomic transition from level $\left|
2\right\rangle $ to $\left| 3\right\rangle $ needs to absorb a $b$-mode
photon. Hence, our scheme is equal to a coupled-cavity waveguide coupled to
a two-level system when $n=0$. However, when $n\neq 0$, the transition $%
\left| 2\right\rangle \leftrightarrow \left| 3\right\rangle $ participates
in the dynamic process, revealing differen behaviors from the $n=0$ case due
to the quantum interference between different atomic transitions.

The arbitrary state governed by the Hamiltonian \eqref{Hamiltonian} can be
written as%
\begin{equation}
\left| \Psi \right\rangle =\sum_{j}\alpha _{j}a_{j}^{\dagger }\left|
1,n\right\rangle \left| \phi \right\rangle +\beta \left| 2,n\right\rangle
\left| \phi \right\rangle +\zeta \left| 3,n-1\right\rangle \left| \phi
\right\rangle \text{,}  \label{state}
\end{equation}%
with $\alpha _{j}$, $\beta $ and $\zeta $ probability amplitudes. The state $%
\left| m,n\right\rangle $, denotes that the atom is in the state $\left|
m\right\rangle $, and the two-mode cavity contains $n$ $b$-mode photons. The
state $\left| \phi \right\rangle $ denotes the 1D waveguide does not contain
any $a_{j}$-mode photon. From the eigenequation $H\left| \Psi \right\rangle
=E\left| \Psi \right\rangle $, we can obtain a set of equations of the
probability amplitudes. Then by eliminating the probability amplitudes $%
\beta $ and $\zeta $, we can obtain the equation of the probability
amplitude $\alpha _{j}$, which reveals the single photon transport property,
as%
\begin{equation}
\lbrack E-(\omega _{a}+n\omega _{b})-V\delta _{j,0}]\alpha _{j}=-\xi (\alpha
_{j+1}+\alpha _{j-1})\text{,}  \label{field}
\end{equation}%
with
\begin{equation*}
V=\frac{g_{a}^{2}\{E-[\omega _{3}+(n-1)\omega _{b}]\}}{[E-(\omega
_{2}+n\omega _{b})]\{E-[\omega _{3}+(n-1)\omega _{b}]\}-g_{b}^{2}n}\text{.}
\end{equation*}%
The effective potential $V$ resulting from the atom-cavity interaction
located at site $j=0$ modifies the single-photon transport property. If the
two-mode cavity does not contain $b$-mode photons, we find $V=\frac{g_{a}^{2}%
}{E-\omega _{2}}$, in line with the outcome in Ref. \cite{Lan}. We consider
the simple case that the atomic transition $\left| 1\right\rangle
\leftrightarrow \left| 2\right\rangle $ is driven resonantly by the input
photon. The potential is derived as $V\rightarrow \infty $ when $n=0$.
However, if we inject a $b$-mode photon into the two-mode cavity initially,
i.e. $n=1$, the potential is obtained as $V=-\frac{g_{a}^{2}(E-\omega _{3})}{%
g_{b}^{2}}$. Especially, when the $b$-mode photon resonantly drives the
atomic transition, i.e. $\omega _{32}=\omega _{b}$, the potential equals to
zero. Therefore, we can modify the effective potential between $\infty $ and
$0$ by one $b$-mode photon. This implies that the single-photon transport
can be controlled by only one $b$-mode photon.

For the single-photon transport in the 1D waveguide, the spatial dependence
of the amplitude $\alpha _{j}$ can be expressed as%
\begin{equation}
\alpha _{j}=(e^{ikj}+re^{-ikj})\theta (-j)+te^{ikj}\theta (j)\text{,}
\label{expression}
\end{equation}%
where $r$ is the reflection amplitude and $t$ is the transmission amplitude.
The Heaviside step function $\theta (x)$ equals to $1$ when $x$ is larger
than $0$, while it equals to $0$ when $x$ is smaller than $0$. From equation %
\eqref{field} and \eqref{expression}, when $\left| j\right| >1$, we can find
$E=n\omega _{b}+\omega _{a}-2\xi \cos k_{a}$, which corresponds to the
dispersion relation derived above. And when $j=0$, we can derive the
expression of the reflection and transmission amplitudes as%
\begin{eqnarray}
r &=&\frac{-g_{a}^{2}(\delta _{a}+\delta _{b})}{2i\xi \lbrack \delta
_{a}(\delta _{a}+\delta _{b})-g_{b}^{2}n]\sin k+g_{a}^{2}(\delta _{a}+\delta
_{b})}\text{,}  \label{randt} \\
t &=&\frac{2i\xi \lbrack \delta _{a}(\delta _{a}+\delta
_{b})-g_{b}^{2}n]\sin k}{2i\xi \lbrack \delta _{a}(\delta _{a}+\delta
_{b})-g_{b}^{2}n]\sin k+g_{a}^{2}(\delta _{a}+\delta _{b})}\text{,}  \notag
\end{eqnarray}%
with the detunings $\delta _{a}=\omega _{2}-\Omega _{k}$, and $\delta
_{b}=\omega _{32}-\omega _{b}$. Here we have employed the continuity
condition $\alpha _{0}^{+}=\alpha _{0}^{-}$. The relation $\left| r\right|
^{2}+\left| t\right| ^{2}=1$ can be easily verified. Obviously, when $\delta
_{a}+\delta _{b}=0$ and $n\neq 0$, the full transmission of the input single
photon is obtained due to the interferences. When $\delta _{a}(\delta
_{a}+\delta _{b})=g_{b}^{2}n$, the input single photon is completely
reflected.

\begin{figure}[tbp]
\includegraphics*[width=4.2cm, height=3cm]{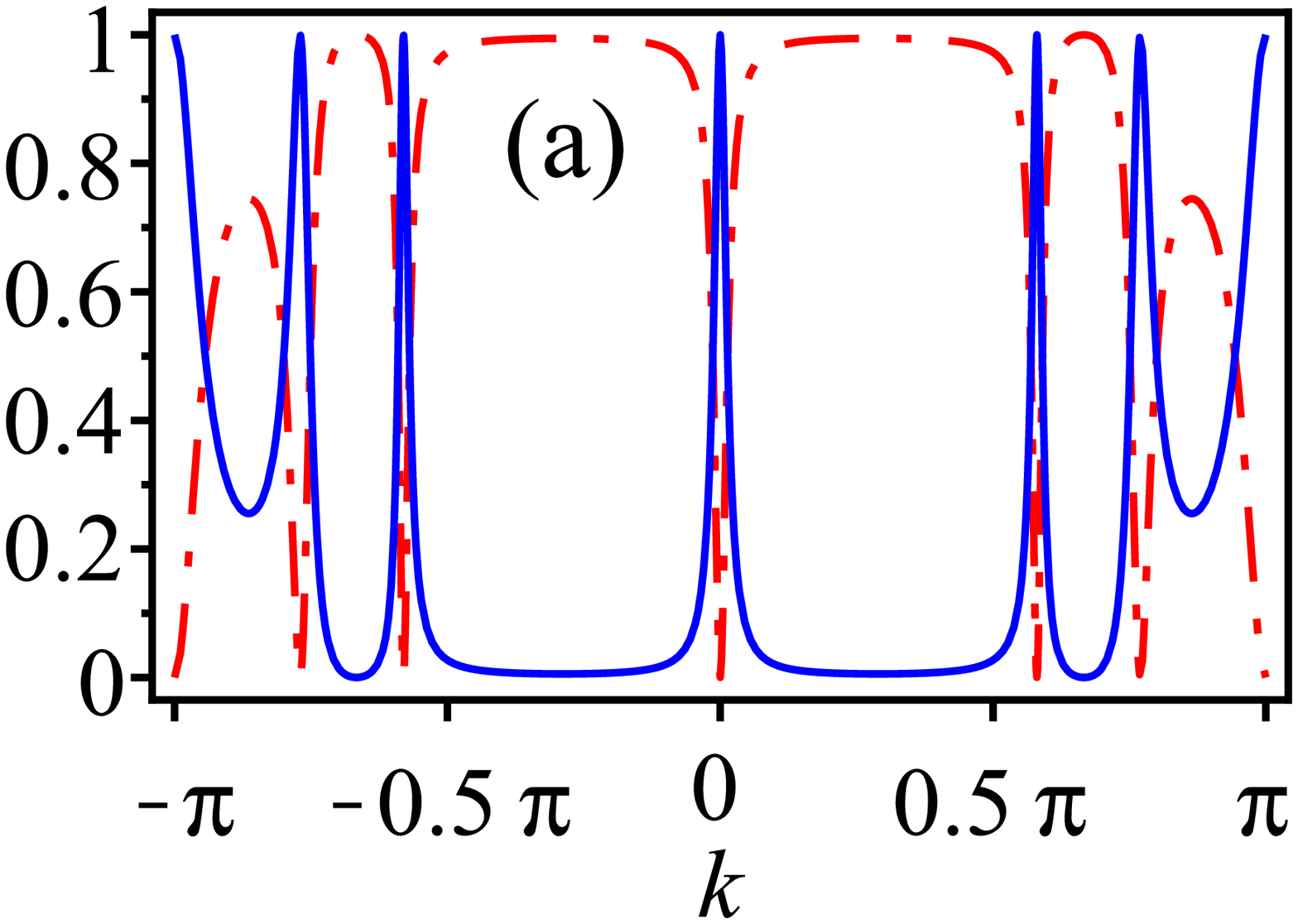} \includegraphics*%
[width=4.2cm, height=3cm]{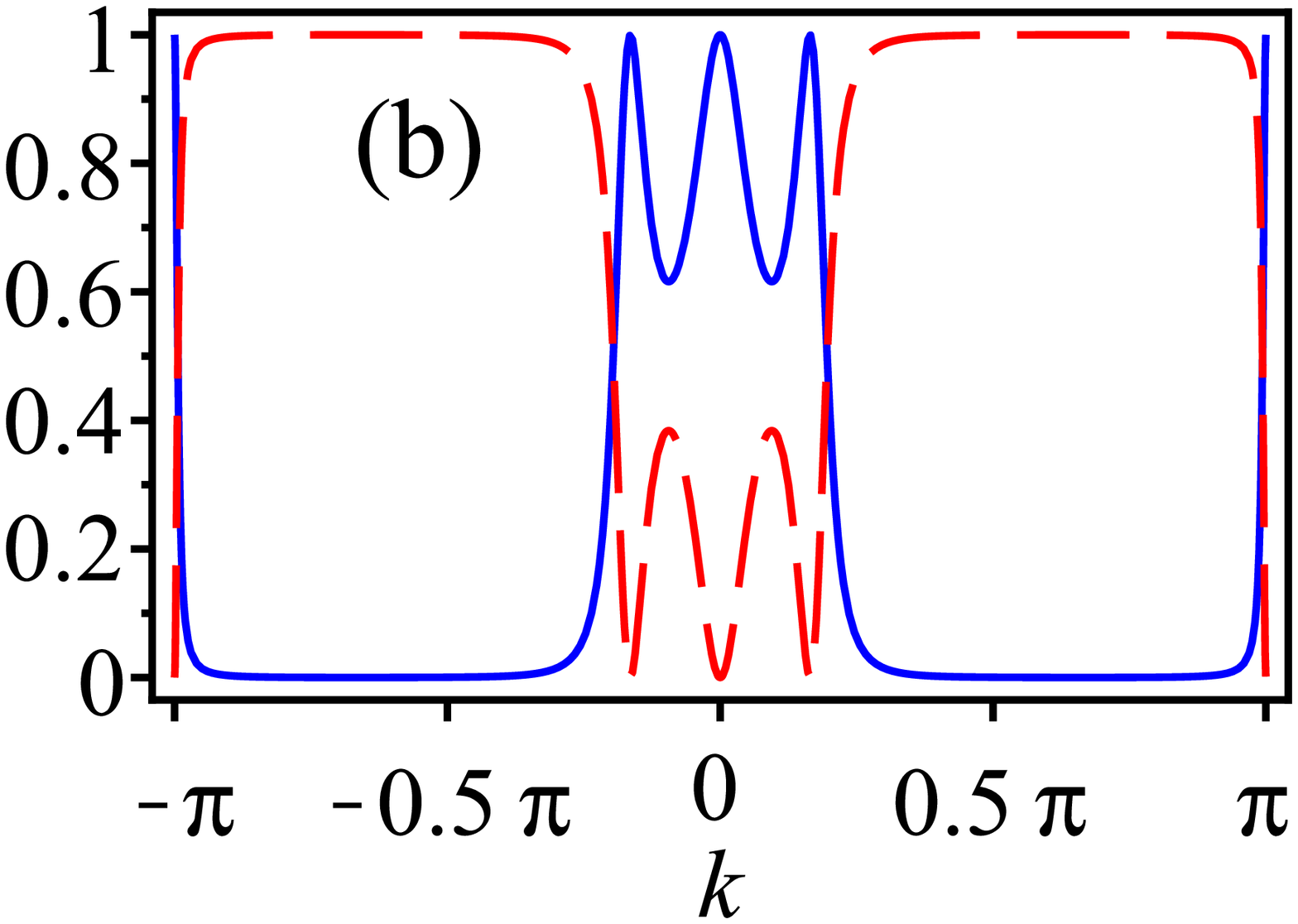} \includegraphics*[width=4.2cm,
height=3cm]{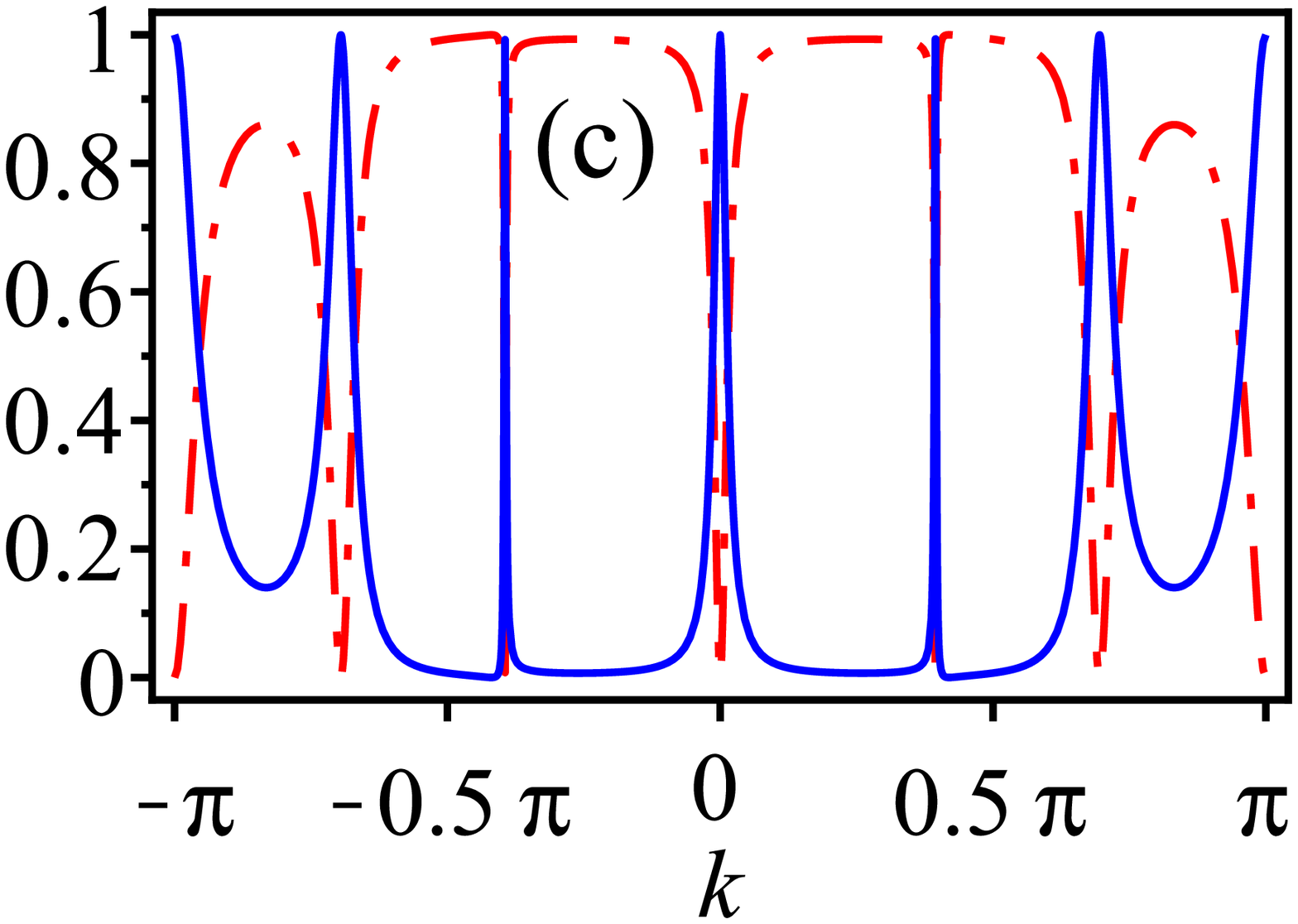} \includegraphics*[width=4.2cm, height=3cm]{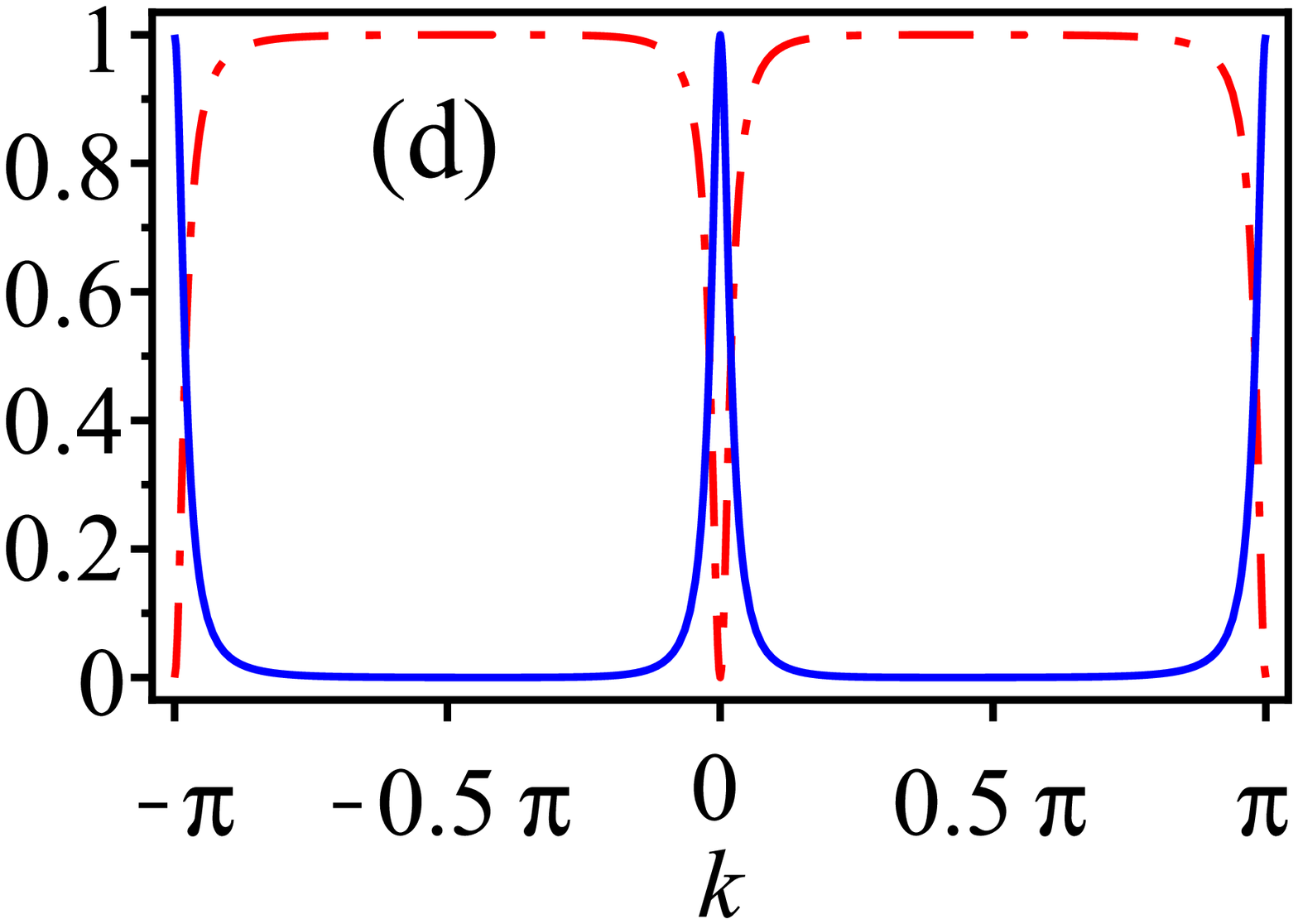} %
\includegraphics*[width=4.2cm, height=3cm]{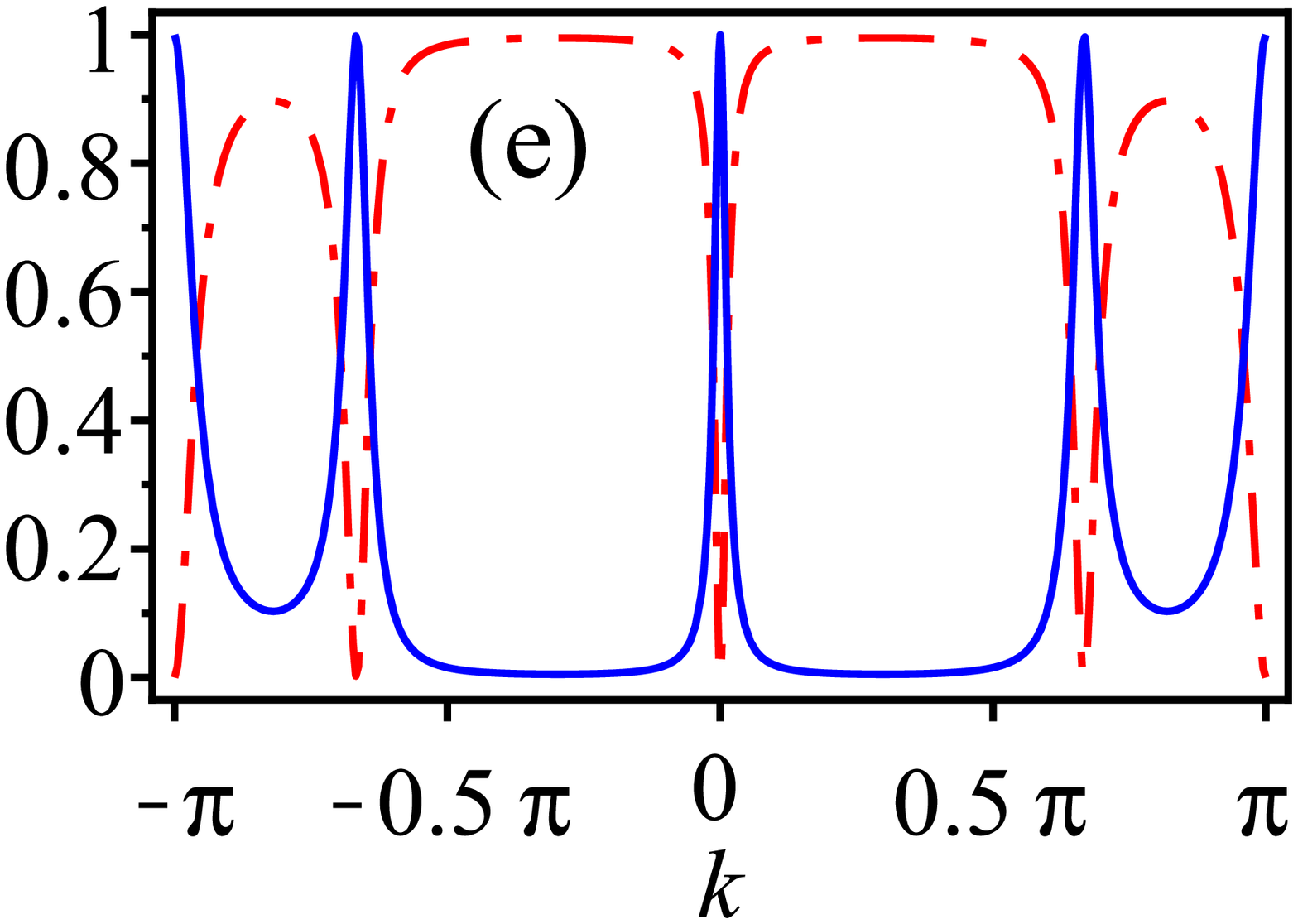}
\caption{The reflection and transmission coefficients against the momentum $%
k $ for various values of $\protect\delta _{b}$ and $n$. The blue solid and
red solid-dot lines represent the reflection and transmission coefficients,
respectively. We take $n=1$ in (a) and (c), and $n=30$ in (b) and (d). In
(a) and (b), $\protect\delta _{b}=0$. In (c) and (d), $\protect\delta %
_{b}=-3 $. We take $n=0$ in (e). The other parameters are $g_{b}=1$, $%
\protect\omega_2-\protect\omega_a=2$, $\protect\xi =2$. All the parameters
but $n$ are in units of $g_{a}$.}
\label{fig2}
\end{figure}

To show the details of the single photon transport property, we plot the
reflection and transmission coefficients as a function of the momentum $k$
in Fig. 2 for various values of $\delta _{b}$ and $n$. The lines show rich
shapes by adjusting\ the $b$-mode photons. The region of $k$ from $-\pi $ to
$\pi $ is enough due to the fact that $r$ and $t$ are periodic functions of $%
k$ with the same period $2\pi $.

From the expression \eqref{randt}, the single-photon transport property
relates to the number of the $b$-mode photons. We first study the resonance
case. When the two-mode cavity does not contain $b$-mode photons, we can
find $r=-\frac{g_{a}^{2}}{2i\xi \sin k\delta _{a}+g_{a}^{2}}$ and $t=\frac{%
2i\xi \sin k\delta _{a}}{2i\xi \sin k\delta _{a}+g_{a}^{2}}$. Obviously,
when $\delta _{a}=0$, the input single photon will be completely reflected.
This outcome has been obtained in a waveguide coupled to a two-level system %
\cite{Lan}. However, when the two-mode cavity contains one or more $b$-mode
photons resonantly driving the atomic transition $\left| 2\right\rangle
\leftrightarrow \left| 3\right\rangle $, i.e. $\delta _{a}=\delta _{b}=0$
and $n>0$, the full transmission is achieved. We note that in the $n>1$
case, the single-photon transport is not affected by the number of $b$-mode
photons in the resonance case. Therefore, the single-photon switch can be
achieved by only one $b$-mode photon.

In the off-resonance case, the single-photon transport relates to the number
of $b$-mode photons even when $n>1$. Here we assume that $g_{a}$ and$\ g_{b}$
have the same order of magnitude. This assumption is reasonable in
experiment. It can be seen that when the number of $b$-mode photons is large
enough, the nearly full transmission of single photon can be obtained. This
can be understood from that $g_{b}\sqrt{n}$ is the effective coupling
strength of the transition $\left| 2,n\right\rangle \leftrightarrow \left|
3,n-1\right\rangle $. When $n$ is large enough, the coupling strength to the
transition $\left| 2\right\rangle \leftrightarrow \left| 3\right\rangle $ is
much larger than the strength to $\left| 1\right\rangle \leftrightarrow
\left| 2\right\rangle $. These behaviors can be understood better in the
dressed representation as shown below.

\begin{figure}[tbp]
\includegraphics*[width=6cm, height=4cm]{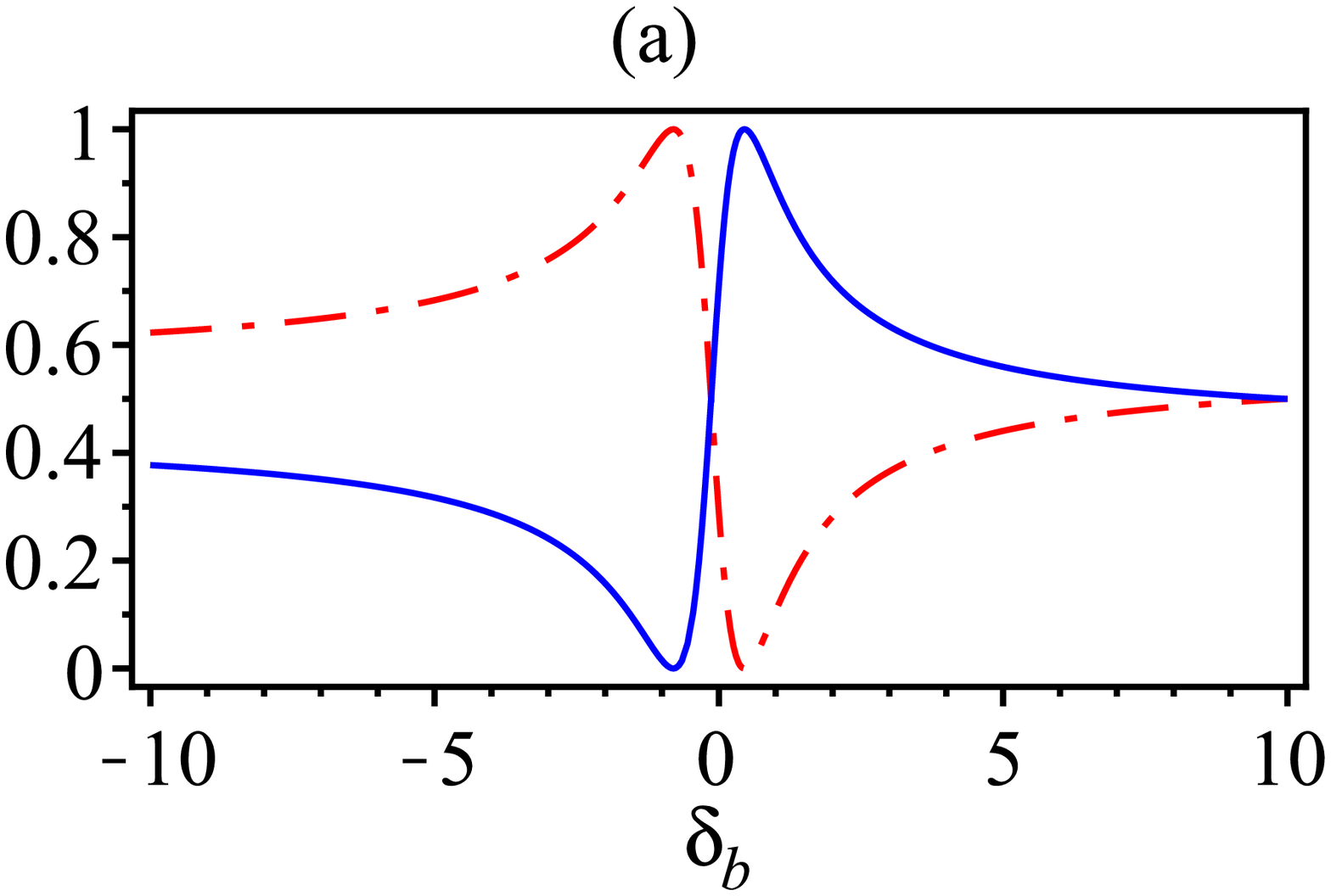} \includegraphics*%
[width=6cm, height=4cm]{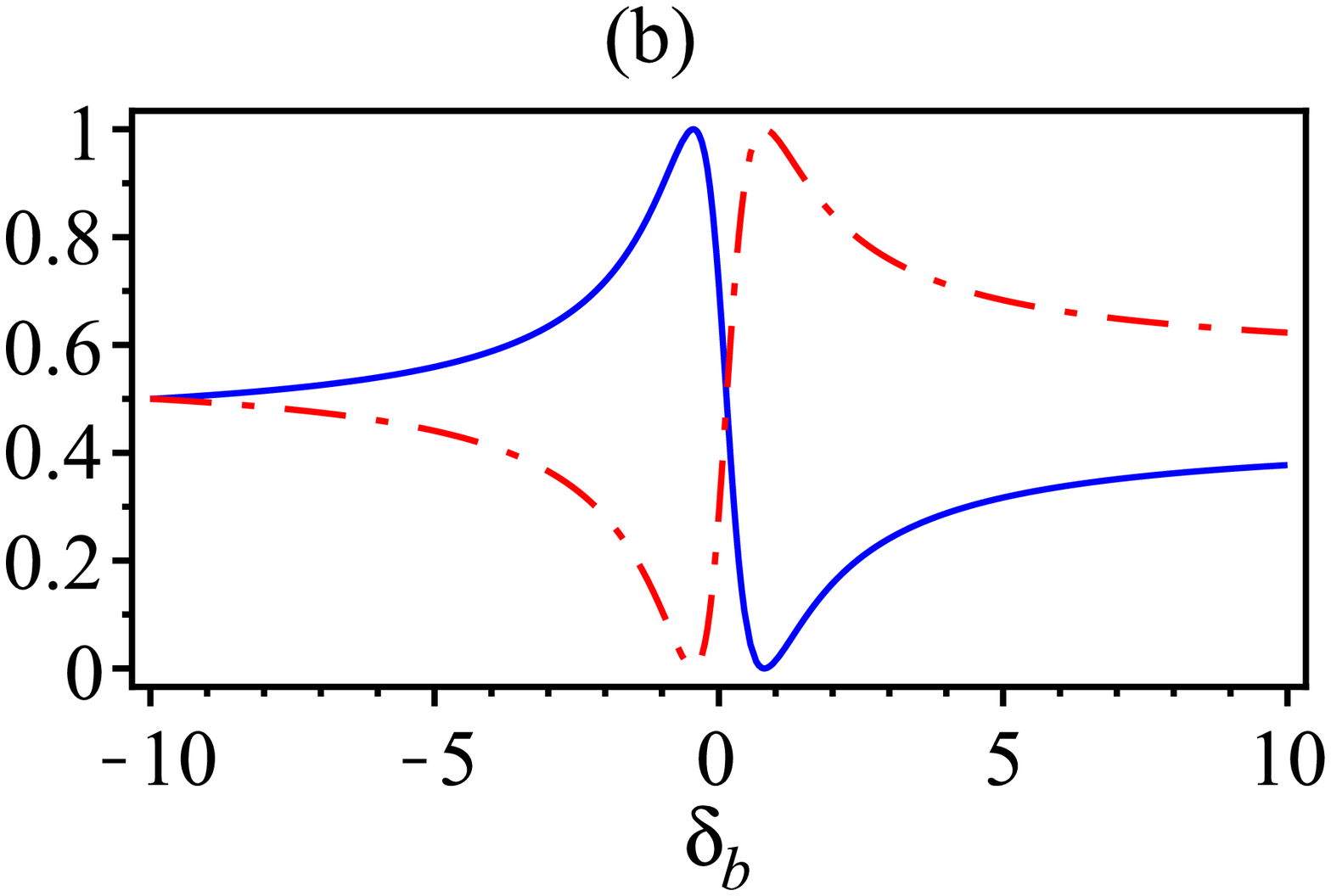} \includegraphics*[width=6cm, height=4cm]{%
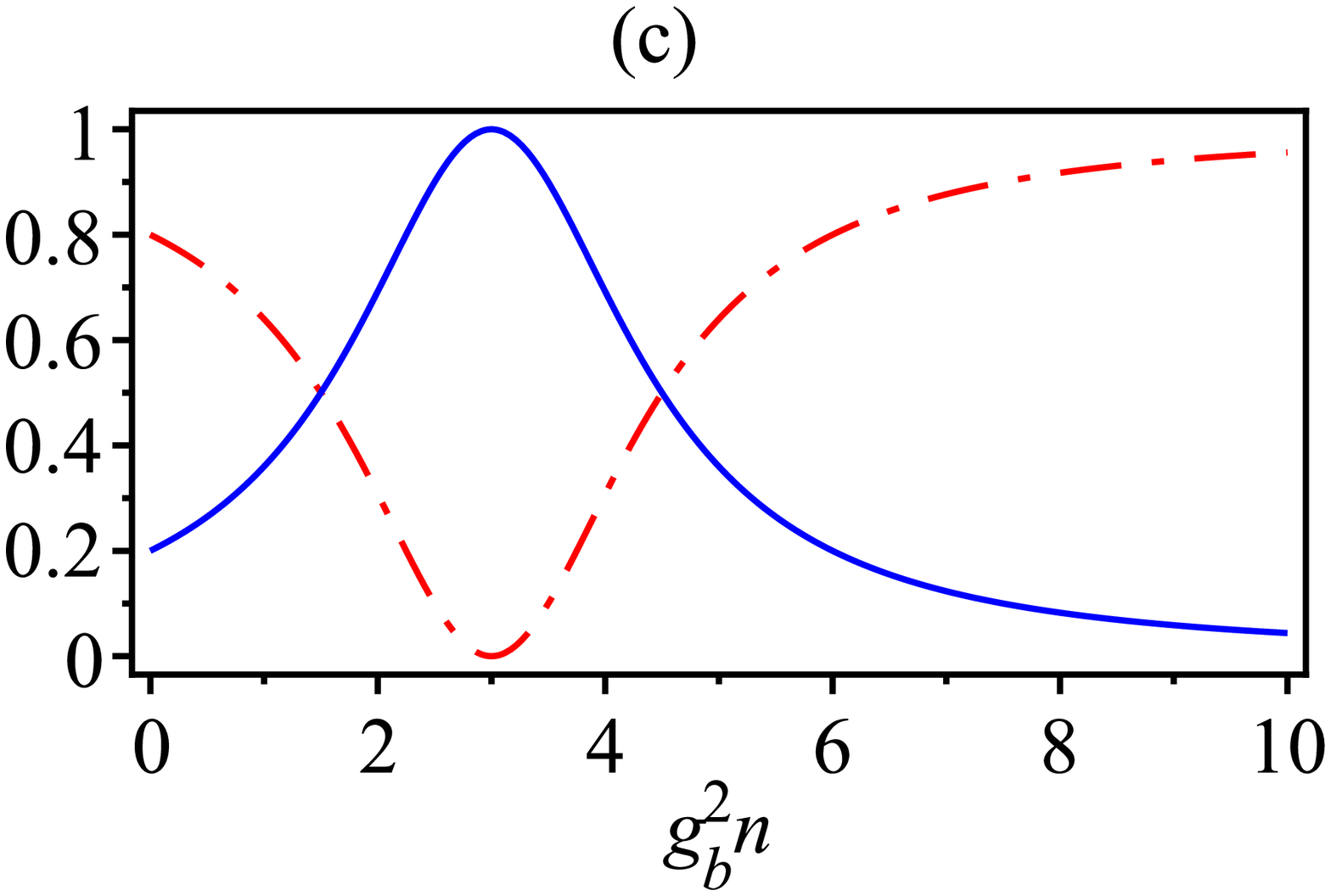}
\caption{The reflection and transmission coefficients against the detuning $%
\protect\delta _{b}$ and $g_{b}^{2}n$. The blue solid and red solid-dot
lines represent the reflection and transmission coefficients,respectively.
(a) and (b) are the coefficients against $\protect\delta _{b}$. We take $%
\protect\delta _{a}=0.8$ in (a) and $\protect\delta _{a}=-0.8$ in (b). The
other parameters of (a) and (b) are $g_{b}=1$, $k=\protect\pi /4$, $\protect%
\xi =1$, $n=1$. (c) is the coefficients against $g_{b}^{2}n$. The parameters
of (c) are $g_b=1$, $k=\protect\pi /6$, $\protect\xi =2$, $\protect\delta %
_{a}=1$ and $\protect\delta _{b}=2$. All the parameters but $n$ are in units
of $g_{a}$.}
\label{fig3}
\end{figure}

Obviously, the single-photon transport property is affected by other
parameters of the $b$-mode cavity, such as the resonant frequency and
atom-cavity coupling strenghs. In Fig. 3a and 3b, we plot the reflection and
transmission coefficients against the detuning $\delta _{b}$ when the
two-mode cavity contains $b$-mode photons and the input photon is
off-resonant to the transition $\left| 1\right\rangle \leftrightarrow \left|
2\right\rangle $. The detuning $\delta _{b}$ can be adjusted by the atomic
transition frequency and the cavity resonant frequency. The control of
single photon transport by tuning the resonant frequency of cavities has
been studied in \cite{Liaojq}. In our scheme, we bring in an extra cavity
which is not in the array of the coupled cavities and is connected with the
array by the three-level emitter. The line shapes are like Fano and
anti-Fano shapes. This is in line with the conditions of full transmission
and reflection obtained below Eqs. \eqref{randt}. Fig. 3c is the the
reflection and transmission coefficients against the parameter $g_{b}^{2}n$.
The feasibly controllable coupling strength between the cavity and the
quantum dot has been proposed in Ref. \cite{Weilf}.

\section{Dressed-state representation}

We can consider that the coupling between the $b$-mode photons and the
atomic transition $\left| 2\right\rangle \leftrightarrow \left|
3\right\rangle $ leads to dressed states. The expression of the dressed
states can be obtained as%
\begin{equation}
\left| \Psi _{\pm }\right\rangle =A_{\pm }\left| 2,n\right\rangle +B_{\pm
}\left| 3,n-1\right\rangle \text{,}  \label{dressed}
\end{equation}%
with the corresponding energies%
\begin{equation}
\omega _{\pm }=\omega _{2}+n\omega _{b}+\frac{\delta _{b}\pm \sqrt{\delta
_{b}^{2}+4g_{b}^{2}n}}{2}\text{.}  \label{omega}
\end{equation}%
Here $A_{\pm }=\frac{-\delta _{b}\pm \sqrt{\delta _{b}^{2}+4g_{b}^{2}n}}{%
\sqrt{2\sqrt{\delta _{b}^{2}+4g_{b}^{2}n}(\sqrt{\delta _{b}^{2}+4g_{b}^{2}n}%
\mp \delta _{b})}}$ and $B_{\pm }=\frac{2g_{b}\sqrt{n}}{\sqrt{2\sqrt{\delta
_{b}^{2}+4g_{b}^{2}n}(\sqrt{\delta _{b}^{2}+4g_{b}^{2}n}\mp \delta _{b})}}$.
Thus, our scheme can be considered as a waveguide coupled to a V-type system
in the dressed-state representation, as shown in Fig. 4. The effective
coupling strength of the $a_{0}$-mode photon to the transition $\left|
1,n\right\rangle \leftrightarrow \left| \Psi _{\pm }\right\rangle $ can be
derived as $g_{\pm }=g_{a}A_{\pm }$. Hence, the Hamiltonian represented in
the dressed-state representation has the form of $H_{C}=\sum_{i=\pm }\omega
_{i}\sigma ^{ii}+\omega _{a}a_{j}^{\dagger }a_{j}-\xi
\sum_{j}(a_{j+1}^{\dagger }a_{j}+h.c.)+\sum_{i=\pm }(g_{i}a_{0}\left| \Psi
_{i}\right\rangle \left\langle 1,n\right| +h.c.)$. Especially, When $n=0$,
we can find $A_{\pm }=1$ and $B\pm =0$, and the effective V-type system
becomes a two-level system.

\begin{figure}[tbp]
\includegraphics*[width=8cm, height=4cm]{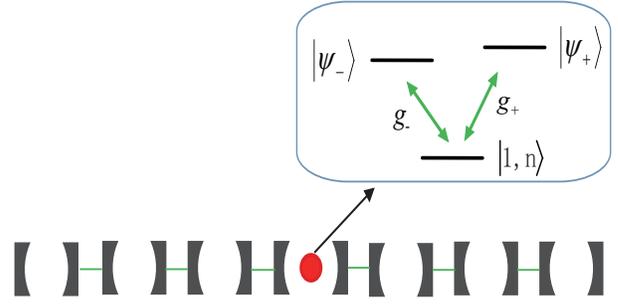}
\caption{In the dressed-state representation, the scheme shown in Fig. 1 can
be considered as a 1D waveguide coupled to an effective V-type atom with
effective coupling strengths $g_{\pm }$.}
\label{fig6}
\end{figure}

It is necessary to bring in two parameters, $\delta _{\pm }=\delta _{a}+%
\frac{\delta _{b}\pm \sqrt{\delta _{b}^{2}+4g_{b}^{2}n}}{2}$, to represent
the detunings between the transition energy $\left| 1,n\right\rangle
\leftrightarrow \left| \Psi _{\pm }\right\rangle $ and the energy $\Omega
_{k}$ of the incident photon. When one of the two detunings $\delta _{\pm }$
is zero, we can find the potential $V\rightarrow \infty $, and the amplitude
$t=0$. Therefore, for the single-photon transport in a coupled-cavity
waveguide coupled to a V-type system, once one of the atomic transitions
resonantly matches the photon, the input photon will be completely
reflected. We note that the condition $\delta _{\pm }=0$ is equivalent to
the condition $\delta _{a}(\delta _{a}+\delta _{b})=g_{b}^{2}n$ derived in
the bear-state representation.

When $\frac{\delta _{+}}{\delta _{-}}=\frac{g_{+}^{2}}{g_{-}^{2}}$, we can
find $\delta _{a}+\delta _{b}=0$. As mentioned above, the full transmission
can be obtained in this case. Therefore, the full transmission of the
injected single photon can be achieved when $\frac{\delta _{+}}{\delta _{-}}=%
\frac{g_{+}^{2}}{g_{-}^{2}}$ in a coupled-cavity waveguide coupled to a
V-type atom. All these outcomes can also be verified by obaining the
single-photon transport property of a waveguide with nonlinear dispersion
relation coupled to a V-type atom as we have done in the bear-state
representation. These behaviors are similar to the single-photon transport
in a linear waveguide coupled to a V-type atom \cite{Witthaut2010njp}.

The effective energies $\omega _{\pm }$ of the dressed state relate to the
number of the $b$-mode photons. When the value of $n$ is large enough, the
detunings $\delta _{\pm }$ become much larger than the effective coupling
strengths $g_{\pm }$. In this case, the single-photon transmission efficient
is approximately unit because the V-type system is nearly decoupled to the
input single photon.

\section{Conclusions}

The transport of the injected single photon in a 1D waveguide coupled to a
three-level emitter with cascade configuration has been investigated. The
atomic transition from $\left| 2\right\rangle $ to $\left| 3\right\rangle $
is driven by $b$-mode photons. When the emitter is in the ground state
initially, whether the $b$-mode cavity contains photons determines the
effective configuration of the emitter. Consequently, the single-photon
transport property depends on whether the $b$-mode cavity contains a photon
or not. The single-photon transport can also be controlled by the parameters
of the $b$-mode cavity. We also study the dressed-state representation. All
the outcomes are obtained in the strong coupling regime. Our scheme
represents an all-optical device operated at single-photon level.

\section{Acknowledgement}

This work is supported by ``973'' program (2010CB922904), grants from
Chinese Academy of Sciences, NSFC (11175248).

\end{document}